# Comment on "Pi in the Sky" by Frolop and Scott: They Lie

Adrian L. Melott, Dept. of Physics and Astronomy, University of Kansas

**Abstract:**

Frolop and Scott (2016) claim significant 1-1 correspondence between anomalies in the cosmic microwave background (CMB) and the digits of pi, which they call "Pi in the Sky". They have without attribution republished the famous work of Joe Hill (Hill 1911), who first proposed this idea, then repudiated it.

**Introduction:**

Frolop and Scott (2016), hereafter FroDS, have discussed the relationship between the digits of pi and the anomalies in the CMB. It is obvious that Joe Hill recognized this idea in 1909, and in fact published his results (Hill 1911), which we reproduce below. Clearly he understood that although there are appealing coincidences, they are not to be trusted.

**Documentation:** Lyrics of Pie in the Sky, aka Long-haired Preachers.

By Joe Hill

Long-haired preachers come out every night

To tell you what's wrong and what's right

But when asked how about something to eat

They will answer in voices so sweet:

    You will eat, bye and bye

    In that glorious land above the sky

    Work and pray, live on hay

    You'll get pie in the sky when you die.

    That's a lie

And the starvation army they play

They sing and they clap and they pray

'Till they get all your coin on the drum

Then they'll tell you when you're on the bum:

    You're gonna eat, bye and bye, poor boy

    In that glorious land above the sky, way up high

    Work and pray, live on hay

    You'll get pie in the sky when you die

    Dirty lie

Holy Rollers and jumpers come out

They holler, they jump, Lord, they shout

Give your money to Jesus they say

He will cure all troubles today

    And you will eat, bye and bye,

    In that glorious land above the sky, way up high

    Work and pray, boy, live on hay,

    You'll get pie in the sky when you die.

If you fight hard for children and wife

Try to get something good in this life

You're a sinner and bad man, they tell

When you die you will sure go to hell

    You will eat, bye and bye

    In that glorious land above the sky

    Work and pray, live on hay

    You'll get pie in the sky when you die

Workingmen of all countries, unite

Side by side we for freedom will fight

When this world and its wealth we have gained

To the grafters we'll sing this refrain:

    Well, you will eat, bye and bye

    When you've learned how to cook and to fry

    Chop some wood, it'll do you good

    You will eat in the sweet bye and bye

    Yes you'll eat, bye and bye

    In that glorious land above the sky, way up high

    Work and pray, and live on hay

    You'll get pie in the sky when you die

    That's a lie....

## References

Frolop, A., & Scott, D. 2016 arXiv:1603.09703 [astro-ph.CO]

Hill, J. 1911 in Songs of the Industrial Workers of the World (aka The Little Red Songbook), 5th Edition.